\documentclass[smallcondensed]{svjour3}
\smartqed
\usepackage{cite}
\usepackage{amsmath,amssymb,amsfonts,dsfont}
\usepackage{algorithmic}
\usepackage{times}
\usepackage{moreverb}
\usepackage{mathrsfs}
\usepackage{bm}
\usepackage{url}
\usepackage{makecell}
\usepackage[T1]{fontenc}
\usepackage{graphicx}
\usepackage{hyperref}
\usepackage{textcomp}
\usepackage{xcolor}
\usepackage{pifont}
\usepackage{textpos}
\usepackage{multirow}

\def\BibTeX{{\rm B\kern-.05em{\sc i\kern-.025em b}\kern-.08em
    T\kern-.1667em\lower.7ex\hbox{E}\kern-.125emX}}

\usepackage{subcaption}
\newcommand{\chen}[1]{\textcolor{black}{#1}}

\begin{document}

% \title{When Trust Meets Decentralization: A Comprehensive Survey on Decentralized Federated Learning}

\title{Enhancing Trust and Privacy in Distributed Networks: A Comprehensive Survey on Blockchain-based Federated Learning}

\author{Ji Liu\footnotemark[1]\footnotemark[4]\footnotemark[5],
        Chunlu Chen\footnotemark[4]\footnotemark[6],
        Yu Li\footnotemark[7],
        Lin Sun\footnotemark[8],
        Yulun Song\footnotemark[8],
        Jingbo Zhou\footnotemark[7],
        Bo Jing\footnotemark[7],
        Dejing Dou\footnotemark[9]
}

\date{Accepted: March 28, 2024}

\institute{
\footnotemark[1] Corresponding author. \\
\footnotemark[4] Equal contribution. \\
\footnotemark[5] Hithink RoyalFlush Information Network Co. Ltd., Hangzhou, China.\\
\footnotemark[6] Information Science and Electrical Engineering Department, Kyushu University, Fukuoka, Japan.\\
\footnotemark[7] Baidu Inc., Beijing, China.\\
\footnotemark[8] Unicom Digital Tech., Beijing, China.\\
\footnotemark[9] Boston Consulting Group, Beijing, China.
}

\authorrunning{Liu and Chen et al.}

\maketitle

\begin{textblock*}{8cm}(3cm,15cm) % {block width} (coords) 
   {\huge To appear in KAIS}
\end{textblock*}

\begin{abstract}
\chen{While centralized servers pose a risk of being a single point of failure, decentralized approaches like blockchain offer a compelling solution by implementing a consensus mechanism among multiple entities. Merging distributed computing with cryptographic techniques, decentralized technologies introduce a novel computing paradigm. Blockchain ensures secure, transparent, and tamper-proof data management by validating and recording transactions via consensus across network nodes. Federated Learning (FL), as a distributed machine learning framework, enables participants to collaboratively train models while safeguarding data privacy by avoiding direct raw data exchange. Despite the growing interest in decentralized methods, their application in FL remains underexplored. This paper presents a thorough investigation into Blockchain-based FL (BCFL), spotlighting the synergy between blockchain's security features and FL's privacy-preserving model training capabilities.}
% First, we present the taxonomy of BCFL from five aspects, including ****, ***, ***, ***, ***.
First, we present the taxonomy of BCFL from three aspects, including decentralized, separate networks, and reputation-based architectures. 
% Then, we summarize existing methods in ***. 
Then, we summarize the general architecture of BCFL systems, providing a comprehensive perspective on FL architectures informed by blockchain.
% Afterward, we analyze the application of BCFL in *** areas. 
Afterward, we analyze the application of BCFL in healthcare, IoT, and other privacy-sensitive areas.
Finally, we identify future research directions of BCFL. 
\end{abstract}

\section{Introduction}

Blockchain is an innovative technology, which fundamentally reshapes transactions and interplays with various entities such as institutions and governments, and verifies authenticity processes. Initially devised for the digital currency Bitcoin, the capabilities of blockchain transcend its initial purpose, providing support for a diverse range of applications, from Peer-to-Peer (P2P) payment services to the management of supply chains. Essentially, a blockchain functions like a traditional ledger, documenting transactions that involve the transfer of money, goods, or secure data. Its structure, making it virtually impossible to alter data without detection by other users, enhances its security. This characteristic shifts verification systems from centralized to decentralized, where the consensus of multiple users facilitates the validation process.

The mechanics of blockchain involve the aggregation and organization of data into blocks, subsequently fortifying these blocks through cryptography. At the core of blockchain security lies a hash function, a cryptographic algorithm that solidifies the connections between blocks. This hash function generates a unique character string for each block, intricately interwoven into the succeeding block, forming a secure chain. Any endeavors to tamper with a previously established block disrupt this chain of hashes, resulting in a mismatch and exposing the attempted modification. The adaptability of blockchain extends beyond tracking commercial transactions; it can effectively store and safeguard sensitive information. Despite its vast potential, this technology is still in its infancy and must overcome various obstacles before achieving widespread adoption. Nevertheless, blockchain signifies a revolutionary shift in how entities and individuals engage, presenting a straightforward and secure mechanism to establish trust for virtually any transaction type.

Federated Learning (FL) has been a very hot topic in recent years. First, as people pay more attention to data privacy, more and more users are unwilling to share their private data. In addition, various countries have also introduced corresponding laws and regulations to restrict the behavior of data collectors, such as the Cybersecurity Law of the People’s Republic (CLPR) of China \cite{CCL}, the General Data Protection Regulation (GDPR) \cite{GDPR}, the California Consumer Privacy Act (CCPA) \cite{CCPA}, and the Consumer Privacy Bill of Rights (CPBR) \cite{Gaff2014}, Internet companies need to be further responsible for user data.FL is a machine learning technique in which people train algorithms on multiple distributed edge devices or servers with local data samples. This approach differs significantly from traditional centralized machine learning techniques, which upload all local datasets to a single server, while more classic decentralized approaches typically assume that the local data samples are all the same Distribution. The emergence of FL protects the privacy of user data to a certain extent, and achieves the effect of ``available and invisible'' \cite{liu2022distributed}.

The fusion of blockchain and FL amalgamates the most advantageous features of both technologies, yielding a resilient and efficient system characterized by heightened data privacy and security. The decentralized nature of blockchain ensures equitable rights for all network nodes, mitigating the risks associated with centralized systems and protecting against data breaches. Its immutable and traceable nature provides inherent data integrity, deterring malicious manipulation and fostering trust among participants. Integrating blockchain into FL promotes a secure and privacy-preserving distributed learning environment, where sensitive data remains on edge devices while ensuring system security and stability. This symbiotic relationship between blockchain and FL paves the way for the development of robust and reliable decentralized machine learning systems, with broad applications in various domains.

Currently, there are numerous researchers exploring BlockChain-based FL (BCFL) architectures. A comprehensive review \cite{berdik2021survey} emphasizes the potential and complexities of blockchain as a service within today's information systems, while highlighting the implications of blockchain in various industries. The integration of blockchain with cloud and edge computing paradigms is underscored as being of paramount importance.
The review \cite{qammar2023securing} explores blockchain's potential to enhance FL by eliminating the need for centralized servers, thus solving problems like private information disclosure and high communication costs.
The concept of BCFL is dissected in a study \cite{li2022blockchain}, with a focus on the unique challenges, structural design, platforms, incentive mechanisms, and applications it presents. A decentralized approach to FL, involving the use of Swarm Learning (SL), is meticulously explored in another investigation \cite{han2022demystifying}. Here, a permissioned blockchain is introduced to ensure secure member onboarding and dynamic leader election, thereby facilitating highly decentralized deep learning.
A blockchain-based solution aimed at enhancing accountability and fairness in FL systems is presented in a different study \cite{lo2021blockchain}. This approach includes a smart contract-based data-model provenance registry and a weighted fair data sampler algorithm. The introduction of SPDL, a decentralized learning scheme integrating blockchain, Byzantine Fault-Tolerant (BFT) consensus, BFT Gradients Aggregation Rule (GAR), and differential privacy, is discussed in another work \cite{xu2022spdl}. This method ensures efficient machine learning while safeguarding data privacy, Byzantine fault tolerance, transparency, and traceability.
Lastly, the Blockchain Assisted Decentralized Federated Learning (BLADE-FL) framework \cite{ma2020federated} is brought forward as a fully decentralized framework, wherein both training and mining responsibilities are assigned to full nodes.
\chen{Numerous studies have explored the architectures of BCFL, shedding light on its potential benefits and inherent challenges. Diverging from earlier efforts that have investigated distinct aspects of BCFL, our paper's value is in amalgamating these diverse methodologies. We aim to offer a unified, exhaustive analysis of BCFL systems. This synthesis not only illuminates the current landscape of BCFL but also lays the groundwork for future advancements in this area. By developing a generic BCFL system architecture that encompasses essential components such as the infrastructure, network, communication, algorithms, blockchain consensus, and application layers, we provide a foundational framework for both evaluating existing FL systems and guiding the development of forthcoming BCFL solutions. Our comprehensive approach aims to catalyze innovation within BCFL by addressing gaps in existing research and proposing new directions for exploration.}
In this paper, we conduct an extensive analysis of BCFL systems. The essence of our work lies in synthesizing existing methodologies and outlining future research directions. A significant contribution of this research is the development of a generic BCFL system architecture. This architecture, structured into key layers including infrastructure, network, communication, algorithms, blockchain consensus, and application, serves as both an evaluative tool for existing FL systems and a foundational guide for the development of future BCFL systems. Moreover, we delve into the diverse applications of BCFL across various sectors, highlighting its versatility and potential for innovation. Our study aims to provide a comprehensive perspective on BCFL systems, while providing the basis for future research in the field of BCFL systems.

The rest of the manuscript is organized as follows. In Section \ref{sec:Overview}, we introduce the basic concepts of blockchain and FL and discuss their relationship. In Section \ref{sec:Taxonomy}, we provide detailed information on the architecture of FL systems, including the infrastructure layer, network layer, communication layer, algorithm layer, Blockchain consensus, and application layer. In addition, we discuss the current state of research and challenges related to each topic. Finally, section \ref{sec:Challenges} and section \ref{sec:Conclusion} discuss the future directions and concludes.

\section{An Overview of Blockchain and Federated Learning}
\label{sec:Overview}

In this section, we introduce the basic concepts of blockchain and FL. Then, we discuss the privacy preserving in FL. And the necessity of integrating blockchain and FL.

\subsection{Blockchain and Distributed System}

\chen{Distributed systems are networks of computers or nodes that work together to form a unified system, characterized by decentralization, consensus, fault tolerance, and scalability.} Decentralization distributes control and decision-making across various nodes, thereby bolstering fault tolerance, scalability, and resilience. Consensus mechanisms protect the system by ensuring that all nodes consistently concur on the state of the system, even under adverse conditions such as failure or malicious actions. Fault tolerance denotes the ability of a distributed system to handle the malfunction of an individual node or component without compromising the functionality of the entire system. This resilience is achieved through strategies such as redundancy, replication, and comprehensive error-handling mechanisms. Scalability refers to the system's capability to horizontally scale by incorporating additional nodes to accommodate increasing workloads and expanding user bases.

\begin{figure*}[htbp]
    \centering
    \includegraphics[width=0.85\textwidth]{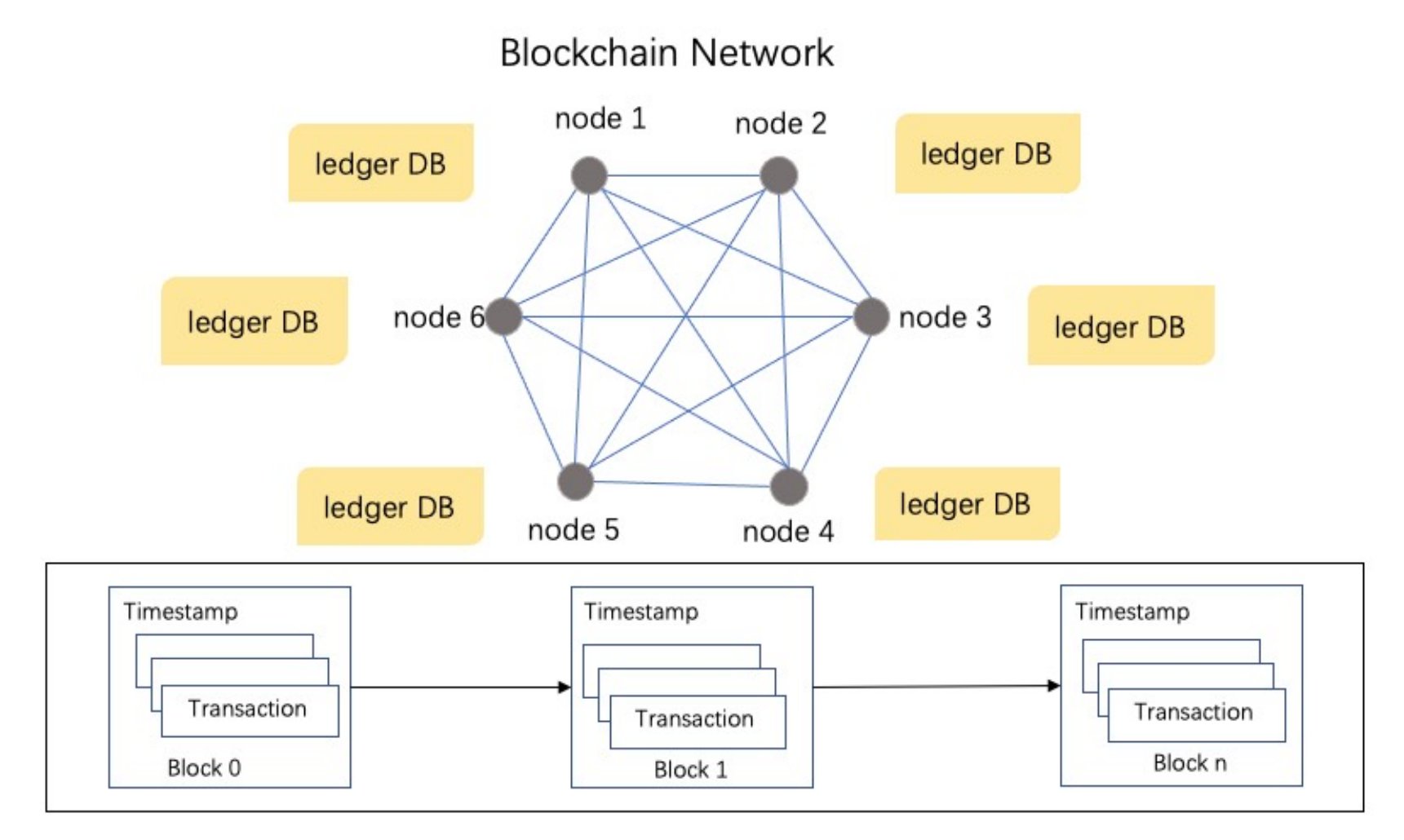}
    \caption{Blockchain Network and Transaction Blocks.}
    %\vspace{-9mm}
    \label{fig:BlockchainNet}
\end{figure*}

Blockchain is a special distributed ledger technology that ensures a secure \cite{kolb2020core,kalodner2020blocksci}, immutable record of transactions or data. Blockchain functions on top of a distributed system and incorporates principles such as cryptography, transparency and immutability, consensus mechanisms, smart contracts, and trust and security, as shown in Figure \ref{fig:BlockchainNet}. Blockchain employs cryptography to safeguard the integrity and security of data, authenticate transactions, and shield information stored on the blockchain through hash functions, digital signatures, and encryption \cite{liang2020secure}. Each transaction recorded on the blockchain is transparent, visible to all network participants, and virtually indelible once added to the blockchain - assuring immutability and traceability. Through a consensus mechanism, blockchain facilitates agreement among distributed nodes on the validity and sequencing of transactions \cite{garay2020sok,wang2019survey}. Supported by most blockchain platforms, smart contracts are self-executing agreements with predefined rules inscribed on the blockchain. These smart contracts autonomously carry out transactions and enforce mutually agreed terms, thereby negating the need for intermediaries. Blockchain technology bolsters trust by eliminating central institutions or intermediaries and makes it challenging for malign actors to tamper with data, courtesy of its distributed features and cryptographic security measures \cite{lugan2019secure,chen2018machine}. In addition, researchers have focused on improving the throughput of blockchain systems \cite{peng2020falcondb}. A method involving the construction of a blockchain system with multiple subchains is outlined in \cite{yu2020ohie}. This system facilitates simultaneous mining operations across all subchains, thereby ensuring security, liveliness, and high throughput in the blockchain protocol.
In another approach, the Red Belly Blockchain Consensus (RBBC) is introduced \cite{crain2021red}. This consensus protocol aims to bolster security and achieve high throughput, particularly in scenarios involving a large number of consensus nodes.

The blockchain system is fundamentally a distributed system, and the core challenges of distributed systems revolve around consistency and consensus. Within distributed systems, the terms synchronous and asynchronous carry specific implications. Synchronization entails that each node in the system has an upper limit on clock error, and message transmission must be completed within a specified time; otherwise, it is deemed a failure. Simultaneously, the processing time for each node to handle the message is predetermined. In synchronous systems, the identification of lost messages is relatively straightforward. On the other hand, asynchronous signifies that each node in the system may have a significant clock difference, and the time taken to process a message at each point can vary arbitrarily, making it challenging to determine where a message has not received a response.
In general, blockchain technology is mainly divided into two types: permissionless, exemplified by Bitcoin, and permissioned, with Fabric being a prominent example. In permissionless systems, users can access the network and blocks anonymously without registration. The network is open for anyone to join or exit freely. The public chain is a decentralized blockchain, ensuring transaction security and immutability through cryptographic (asymmetric encryption) algorithms and establishing mutual trust in a network environment with consensus mechanisms. Common consensus mechanisms in public chains include Proof of Work (POW) and Proof of Stake (POS). In contrast, in permissioned blockchains, the prevalent consensus protocols are Kafka, Raft, and PBFT.

Blockchain's versatility is demonstrated through its applications in sectors like healthcare, where it enhances security, privacy, and interoperability in Electronic Health Record (EHR) systems \cite{shi2020applications}. Additionally, blockchain has been studied in the context of cloud computing, addressing security and privacy challenges when outsourcing computational tasks to cloud service providers \cite{shan2018practical}. 
Additionally, Unmanned Aerial Vehicles (UAVs) \cite{pokhrel2021blockchain}, smart city \cite{esposito2021blockchain}, cloud computing \cite{gai2020blockchain},edge computing \cite{xiong2018mobile}, Internet of Vehicles (IoV) \cite{kang2019toward}, these examples demonstrate the broad range of applications for blockchain technology and its ability to address diverse challenges across different industries.

\subsection{Privacy Preserving in Federated Learning}

With the development of database technology and network technology, all kinds of industries have accumulated a large amount of useful data. How to extract valuable knowledge for decision-making from these data has become a top priority. Positioned as a potent data analysis tool, data mining excels in uncovering latent patterns and regularities within data, presenting findings in the form of rules, clusters, decision trees, dependency networks, or other knowledge representations. These insights find applications in diverse areas such as business decision-making, scientific research, and medical investigations.

\begin{figure*}[htbp]
    \centering
    \includegraphics[width=0.85\textwidth]{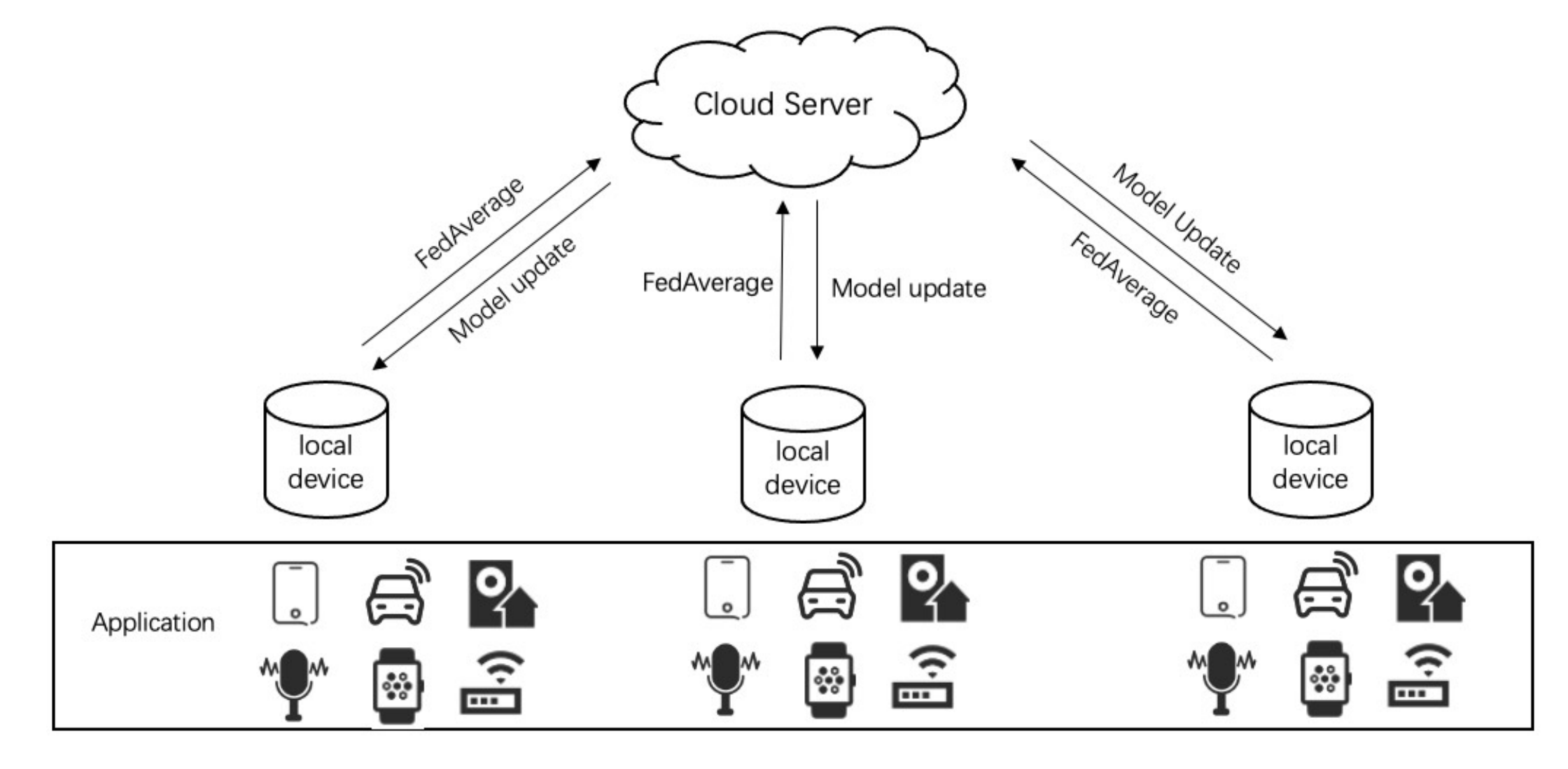}
    \caption{Federated Learning architecture and its applications.}
    %\vspace{-9mm}
    \label{fig:modefl}
\end{figure*}

FL initializes model parameters for all clients via a central server, as shown in Figure \ref{fig:modefl}. The client trains the local model with the initialized model parameters and shares the parameters trained by the local model to the central server. The central server aggregates the parameters of the local model and sends the updated model and parameters to each client. Repeat the above steps until the model converges.

FL is divided into centralized FL and decentralized FL according to the network topology. According to data availability, it can be divided into cross silo FL and cross device FL \cite{li2021survey}. In different data partition scenarios, FL can be divided into horizontal FL, vertical FL and federated transfer learning. Among them, the optimization algorithms of FL include fedavg, smc-avg, fedprox etc.

\begin{itemize}
    \item \textbf{Data Heterogeneity:} Data heterogeneity is a significant challenge in FL due to the decentralized nature of the approach. In FL, data is distributed across multiple devices or entities, each with its unique characteristics, formats, and representations. Firstly, feature heterogeneity arises when participating devices have different sets of features or attributes available for training, making the aggregation and alignment of models challenging. Secondly, data distribution heterogeneity occurs due to variations in user populations, geographical locations, or data collection practices among participating devices, potentially introducing biases into the trained model. Lastly, data format heterogeneity poses challenges in FL as data from different sources or platforms may have varying formats, representations, or structures. 

    \chen{To address these heterogeneity challenges, researchers employ various techniques \cite{ye2023heterogeneous}, including model architecture adjustments and feature engineering to harmonize feature spaces across devices. They utilize meta-learning and domain adaptation to manage data distribution variations and explore models with intermediate representations or multi-format designs for data format heterogeneity. These strategies facilitate effective data use in federated learning, enhancing model robustness and efficiency across diverse environments.}
    \item \textbf{Privacy and Security:}
    FL is a decentralized approach to machine learning where multiple devices or entities collaboratively train a shared model without sharing their raw data. While FL offers numerous advantages, such as preserving data privacy and reducing communication costs, it also raises important privacy concerns \cite{yin2021comprehensive}.

    \begin{itemize}
        \item \textbf{Data Leakage:} During the training process, models are shared among participating devices or entities. There is a risk of unintentional data leakage if the models contain sensitive information about the training data. Adversaries could potentially reconstruct or infer sensitive data from the shared models.
    
        \item \textbf{Membership Inference Attacks:} In a membership inference attack, adversaries aim to determine whether a specific data point was part of the training dataset. By analyzing the model's responses to queries, adversaries can infer the presence or absence of particular data points, potentially revealing sensitive information about individuals.
    
        \item \textbf{Model Poisoning Attacks:} In a model poisoning attack, adversaries inject malicious data or manipulate their local updates to poison the shared model. This can lead to the model incorporating biased or incorrect information, compromising the privacy of other participants' data.
    \end{itemize}

    \chen{To mitigate these threats, techniques such as Differential Privacy (DP), which adds noise to the data or model updates to obscure individual contributions, and Homomorphic Encryption (HE), allowing computations on encrypted data, are applied. These methods ensure that FL remains resilient against attacks while maintaining data privacy and model integrity.}
    The focus of many research endeavors has been on bolstering security and enhancing privacy protection within the realm of FL. For instance, a scalable production system has been outlined in \cite{bonawitz2019towards}, which is tailored for the mobile device domain, tackling challenges related to device availability, unreliable connections, cross-device coordination, and limited resources.
    A different framework is introduced in \cite{li2021ditto}, which employs regularization terms to amplify the fairness and robustness of personalized FL. In \cite{truex2019hybrid}, differential privacy and homomorphic encryption have been incorporated into FL, aiming to diminish privacy noise and enhance model accuracy via secure multi-party computation.
    The TrustFL scheme \cite{zhang2020enabling} leverages Trusted Execution Environments (TEE) to secure the trustworthy execution of training tasks, offering high-confidence guarantees while preserving efficiency. These various methodologies collectively contribute to enhancing the security and privacy of FL.
    Secure model aggregation and performance optimization also emerge as crucial areas of research. A distributed framework enabling joint association and resource allocation is discussed in \cite{khan2021dispersed}, paving the way for multiple groups to learn a global model. In addition, a model-contrastive FL framework is proposed \cite{li2021model}, which improves local training performance by accounting for the similarity of model representations.
    The framework proposed in \cite{pandey2020crowdsourcing} coordinates multiple mobile clients via an MEC server for parameter aggregation and global model updates. Finally, a Bayesian nonparametric neural network framework for FL \cite{yurochkin2019bayesian}, where the global model is constructed using the Probabilistic Federated Neural Matching (PFNM) method to tackle communication issues. Each of these approaches contributes uniquely to the secure model aggregation and performance optimization in FL.
    \item \textbf{Traceability and Accountability:}
    Traceability and accountability are critical considerations in FL to ensure transparency, integrity, and responsible use of data. 

    \begin{itemize}
        \item \textbf{Traceability} involves model auditing, which tracks and audits the models trained in FL by recording metadata such as model architecture, hyperparameters, and data sources. Data provenance is another aspect of traceability, enabling the tracking of the origin and history of the data used for training, ensuring authenticity, and assessing potential biases. These traceability measures enhance transparency and build trust among participants.
    
        \item \textbf{Accountability} in FL encompasses participant accountability, where participants are expected to adhere to agreed-upon protocols, privacy measures, and security practices. Participants should be held accountable for their actions to maintain the integrity of the FL process. Moreover, maintaining security and trust in FL systems requires participants to implement appropriate security measures, protect data confidentiality, and prevent unauthorized access or malicious activities.
        \end{itemize}
    
    However, traceability and accountability in FL face several challenges. The decentralized structure of FL introduces complexity in coordinating and establishing consensus on traceability standards and accountability mechanisms across multiple participants. Ensuring traceability and accountability while preserving privacy adds another layer of complexity, as privacy-preserving techniques must balance providing traceability information while protecting sensitive participant data. Additionally, data fragmentation caused by the distribution of data across different participants, poses challenges in tracing the origin and lineage of data.

\end{itemize}

\subsection{The Necessity of Integrating Blockchain and Federated Learning}

Combining blockchain and FL has the potential to offer several advantages and address key challenges in both domains:
\begin{itemize}
        \item \textbf{Incentive Mechanisms:} The cryptocurrency or token system native to blockchain can facilitate the establishment of incentive mechanisms for participants in FL. In traditional FL, due to economic rationality, many clients are reluctant to share their valuable data. By rewarding data contributors, model validators, and other participants with tokens, blockchain can incentivize active participation, data sharing, and model improvement \cite{liu2020fedcoin}.

        \item \textbf{Enhanced Data Privacy and Security:} The decentralized and immutable nature of blockchain offers a secure and transparent framework for data sharing and storage. The traditional FL framework heavily relies on a single central server and may fall apart if such a server behaves maliciously. At the same time, the existing design is vulnerable to the malicious clients that might upload poisonous models to attack the FL network. \chen{Incorporating the cryptographic methods of blockchain into Federated Learning (FL) significantly bolsters data privacy and security. Blockchain's decentralized and immutable nature ensures that data remains encrypted and anonymous, safeguarding against the vulnerabilities of centralized servers and malicious model uploads by clients. By leveraging blockchain, FL can achieve a transparent framework for data sharing and storage, enabling participants to validate the integrity and authenticity of shared models without exposing sensitive information \cite{moudoud2021towards, kang2020reliable, li2021local}. This integration enhances trust and reliability within the FL ecosystem.}

        \item \textbf{Trust and Transparency:} The transparency and auditability inherent in blockchain can tackle trust issues in FL. It enables participants to track the history and provenance of data, models, and computations. This transparency fosters trust among participants, as they can verify the fairness and reliability of the FL process \cite{yang2023explainable}.
        
\end{itemize}

The application of Blockchain in FL is exemplified by various frameworks and systems designed to enhance privacy, accuracy, and trust:
\begin{itemize}
    \item BML-ES \cite{tian2021blockchain}: A blockchain-centric machine learning framework for Industrial Internet of Things (IIoT) edge services, utilizing smart contracts for aggregation strategies and employing the SM2 public key cryptosystem to secure privacy and improve model accuracy.
    
    \item TrustFed \cite{ur2021trustfed}: Integrates blockchain in cross-device FL systems to prevent model poisoning, ensure fair training, and maintain the reputation of participants. Smart contracts are used to manage reputations and exclude malicious actors, ensuring a dependable training environment.
    
    \item State Channels for Trust Supervision \cite{zhang2021federated}: A mechanism that uses blockchain and FL to create a trusted environment for distributed data sharing. It employs state channels to establish secure sandboxes for FL tasks, ensuring integrity and supervision throughout the process.
    
    \item Proof of Federated Training (PoFT) \cite{chakraborty2022proof}: A framework enabling verifiable model training across blockchain networks, enhancing transparency and trust in the collaborative training process.
    
    \item Decentralized Model Training and Gradient Aggregation \cite{zhao2021blockchain}: Proposes a blockchain-based architecture for secure model training and introduces a gradient aggregation method aimed at enhancing model accuracy, privacy, and performance.
    
    \item Blockchain-in-the-loop FL \cite{mothukuri2021fabricfl}: Merges traditional FL with Hyperledger Fabric, incorporating gamification to enhance participation and efficiency.
\end{itemize}
\chen{Smart contracts within these systems play a pivotal role in ensuring fairness and dependability by automating enforcement of agreements and conditions without the need for intermediaries. This automation ensures that all parties adhere to the predefined rules, significantly reducing the risk of biased or malicious behavior. As a result, smart contracts contribute to creating a transparent, secure, and trustworthy environment for FL, facilitating its application in diverse fields such as smart cities \cite{zheng2022applications}, vehicular communication networking \cite{pokhrel2020decentralized}, edge FL \cite{hu2021blockchainB}, precision medicine \cite{warnat2021swarm,wang2021blockchain}, thereby enhancing privacy, security, and trust in FL ecosystems \cite{Ouyang2023Artificial}.}

\section{Blockchain-based Federated Learning: A Taxonomy and Review}
\label{sec:Taxonomy}
\begin{figure*}[htbp]
    \centering
    \includegraphics[width=0.70\textwidth]{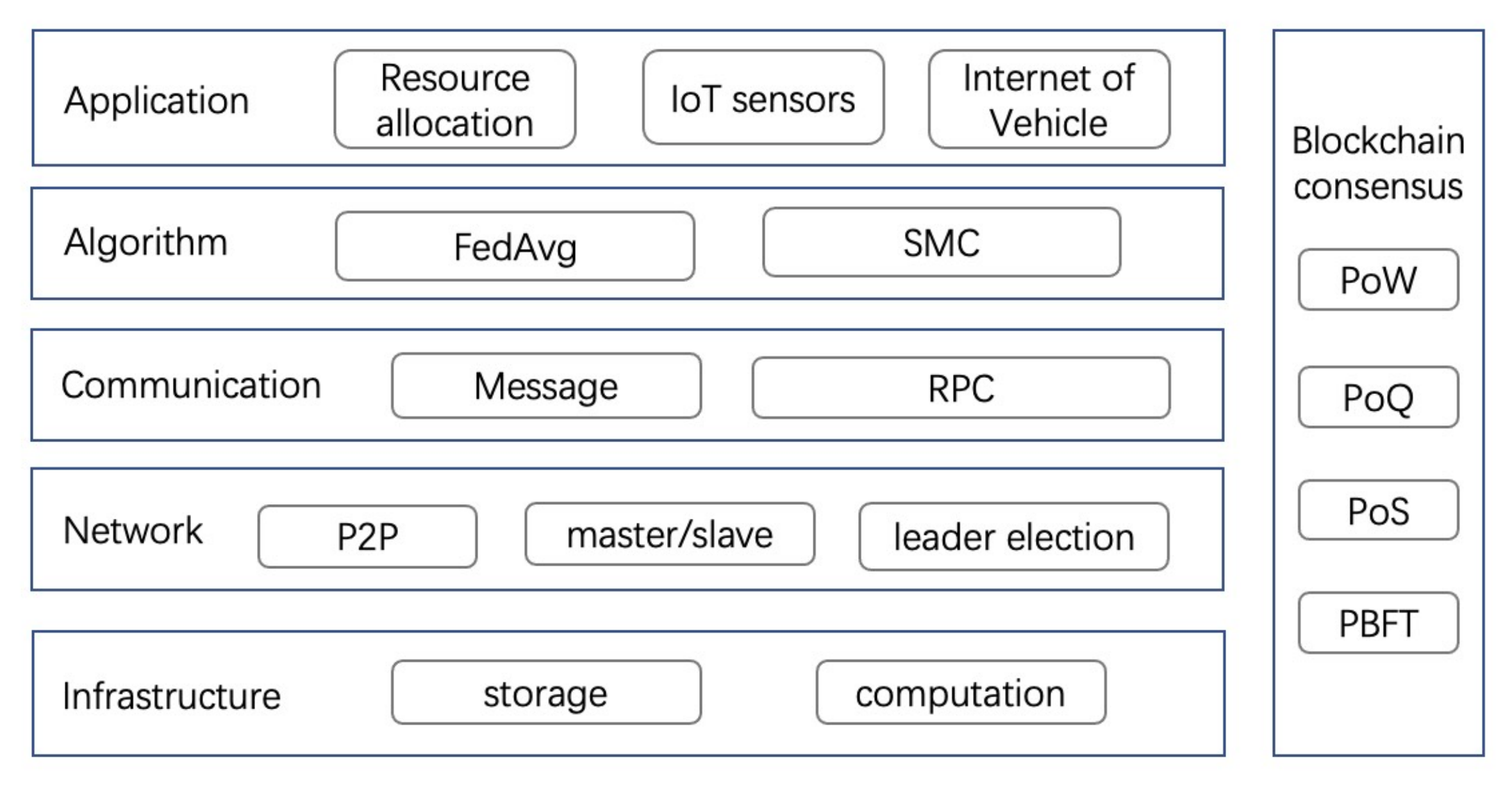}
    \caption{Blockchain-based Federated Learning Architecture.}
    %\vspace{-9mm}
    \label{fig:Bmodelp}
\end{figure*}

From the architecture of the FL in figure \ref{fig:modefl} we can realize that the aggregator server suffers from the single point of failure. However, with the adaptation of blockchain technology into FL, some security assumptions for FL need to be declared in advance. First, aggregator may behave dishonestly or external attacks could influence the FL result. Therefore, numerous works have been done by combining blockchain and FL technology in order to prevent malicious or honest-but-curious aggregator \cite{chen2018machine}. 

Usually, the architecture of this kind of BCFL is shown as figure \ref{fig:Bmodelp}. From a bottom-up perspective, the overall architecture is divided into five layers. The bottom layer of the architecture is infrastructure layer which contains various kind of storage and computation resource. We will discuss and a taxonomy will be provided in section 3.1. Section 3.2 will discuss the network layer of blockchain based FL, which mainly includes P2P, random leader election etc. There are mainly two ways in communication layer which contains message and RPC protocol. In the algorithm layer, privacy preserving methods and incentive mechanism are provided in this layer. Then the upmost layer is related application. Besides, the blockchain consensus plays a vital role in the whole architecture.  

\subsection{Architectures of Blockchain-based Federated Learning}

\chen{BCFL Systems are categorized into three distinct architectures based on their operational dynamics \cite{wang2021blockchain, zhu2023blockchain}: fully coupled, where clients double as both training and blockchain nodes, offering decentralization but requiring high device performance; flexibly coupled, which separates blockchain and FL operations to ease network communication, achieved through committee selection or smart contracts; and loosely coupled, prioritizing reputation to gauge participant reliability, focusing mainly on model update validation and reputation management on the ledger.}

\subsection{Infrastructure Layer}

\chen{The BCFL architecture incorporates FL into blockchain in two primary configurations: the L1 and L2 layer architectures. The L1 layer directly integrates FL with blockchain, offering a decentralized, P2P framework where peers join freely, supported by consensus mechanisms for system reliability \cite{shayan2020biscotti}. The L2 layer, alternatively, builds FL atop blockchain nodes, emphasizing layered data processing and model training \cite{bhattacharya2019bindaas}.}

\chen{At its core, the BCFL infrastructure unites data storage and computational resources. It adopts a decentralized storage model, with privacy ensured through advanced cryptographic methods like SMPC and homomorphic encryption. These technologies facilitate secure, private data sharing and model aggregation, while blockchain's auditability improves transparency and trust. Computational demands are met locally, utilizing Central Processing Unit (CPUs), GPUs \cite{2012cuda}, and Tensor Processing Units (TPUs) \cite{2017tpu}. The system addresses device heterogeneity \cite{che2022federated} and computational limits through lightweight architectures, model compression \cite{jia2023efficient}, and federated distillation, optimizing performance and resource allocation across varied devices.}

\subsection{Network Layer}
\begin{figure}[htbp]
    \centering
    \begin{subfigure}[t]{0.28\textwidth}
      \includegraphics[width=\textwidth]{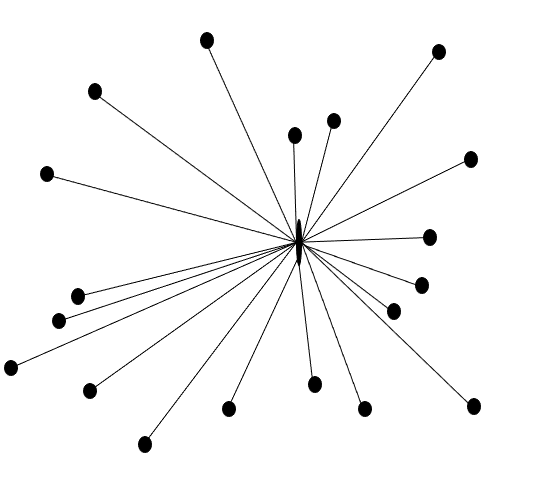}
      \caption{centralized}\label{network:1}
    \end{subfigure}
    \begin{subfigure}[t]{0.28\textwidth}
      \includegraphics[width=\textwidth]{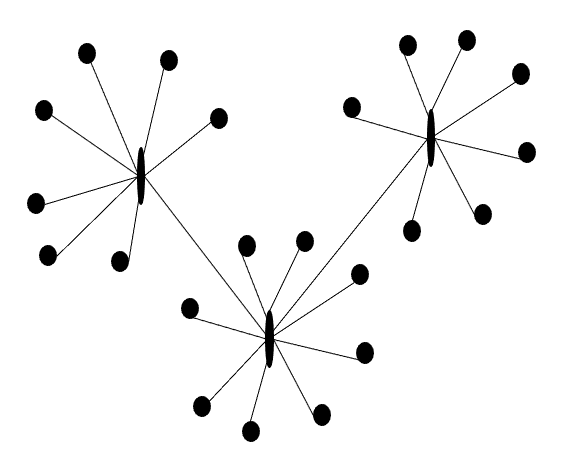}
      \caption{multi-centralized}\label{network:2}
    \end{subfigure}
    \begin{subfigure}[t]{0.23\textwidth}
      \includegraphics[width=\textwidth]{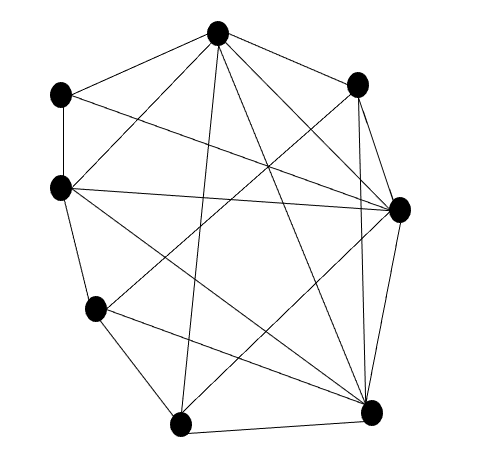}
      \caption{decentralized}\label{network:3}
    \end{subfigure}
    %\vspace{-1.5ex}
    \caption{
    Blockchain network structure categories
    }   \label{fig:network}
    % \vspace{-1ex}
\end{figure}

The network structure of blockchain inherits the general topology structure of computer communication network, and can be divided into three categories: centralized network, multi-centralized network and decentralized network as shown in figure \ref{fig:network}. 

\begin{itemize}
    \item \chen{Centralized Networks: In a centralized network setup as shown in figure \ref{fig:network}(a), all communications and transactions are routed through a central node. This structure, often seen in permissioned blockchain systems, can offer streamlined efficiency and quicker consensus due to the singular control point. However, it also presents a significant security risk; the central node becomes a prime target for attacks, potentially compromising the entire network's integrity and privacy. For BCFL, this setup could limit the system's resilience and increase vulnerability to data breaches and single points of failure.}
    \item \chen{Multi-Centralized Networks: Multi-centralized networks as shown in figure \ref{fig:network}(b), or federated blockchains, introduce several central nodes instead of just one. This setup is typically employed to balance control among multiple organizations or parties, enhancing collaboration while still maintaining a level of centralized governance. While this structure improves security and reduces the risk associated with a single point of failure, it may still face challenges in achieving the same level of decentralization and resistance to censorship or collusion as fully decentralized networks. For BCFL systems, multi-centralized networks can offer improved security and operational efficiency but may still encounter scalability limits and centralized control issues.}
    \item \chen{Decentralized Networks: Decentralized networks as shown in figure \ref{fig:network}(c), epitomized by permissionless blockchains like Bitcoin and Ethereum, distribute data verification and transaction processing across a wide array of nodes. This P2P network structure ensures no single point of control or failure, significantly enhancing security and data integrity. Each node operates with equal status, creating a robust system resistant to censorship, tampering, and attacks. For BCFL, decentralized networks provide a secure and transparent environment for data sharing and model training, although they may face challenges in terms of scalability and consensus speed due to the distributed nature of decision-making.}
\end{itemize}

\chen{In summary, the choice of network setup in BCFL systems profoundly impacts their performance and security. Centralized networks may offer operational efficiency but pose higher security risks. Multi-centralized networks provide a balance with improved security but still retain some centralized control aspects. Decentralized networks, while offering the highest level of security and data integrity, might struggle with scalability and slower consensus mechanisms.}

\subsection{Communication Layer}

The communication layer plays a pivotal role in BCFL systems, orchestrating data transfer and message exchanges among participants to facilitate the collaborative learning process. It is tasked with transmitting model updates, aggregated results, and coordination directives efficiently and securely across the network of devices or servers engaged in FL.

The communication layer employs various protocols and mechanisms to ensure efficient and secure data exchange. 
\begin{itemize}
    \item Encryption Techniques: Advanced encryption methods are utilized to encrypt data before transmission, safeguarding against unauthorized access and ensuring that data privacy is maintained.
    \item Network Protocols: Reliable and efficient network protocols are employed to manage the delivery of messages. These protocols are designed to ensure that data packets reach their intended destinations reliably and in order, even in the face of network disruptions or congestion.
    \item Synchronization Mechanisms: Given the distributed nature of BCFL systems, synchronization mechanisms are crucial for coordinating communication among participants \cite{liu2023distributed}. These mechanisms ensure that model updates are shared in a timely manner and help in managing the asynchronous nature of data transmissions, aligning the updates from various participants.
    \item Decentralized Communication Frameworks: Employing decentralized frameworks, such as peer-to-peer (P2P) networks or blockchain, facilitates direct communication between participants, eliminating the need for centralized intermediaries. This approach not only enhances the system's resilience and decentralization but also reduces potential bottlenecks and points of failure. For example, the Blockchain Assisted Decentralized FL (BLADE-FL) framework is proposed in \cite{ma2020federated}. This fully decentralized framework assigns responsibility for both training and mining to full nodes. These frameworks can enhance the decentralization and resilience of the FL system.
\end{itemize}

\chen{However, BCFL systems face significant challenges such as latency and limited bandwidth, which are addressed through data compression  techniques, to minimize the size of data transmissions and enhance exchange speeds. Adaptive network routing and congestion control algorithms optimize data flows, reducing latency and improving communication efficiency. Additionally, batch processing and caching are employed to lower the frequency and volume of data transfers, mitigating the impact of network constraints. These strategies ensure the communication layer in BCFL systems facilitates secure, efficient, and resilient data exchange, overcoming the hurdles of distributed learning environments and promoting seamless collaboration among network participants.}

\subsection{Algorithm Layer}

In the algorithm layer of BCFL, there are four key aspects to consider: Aggregation algorithms, security algorithms, optimization algorithms, and incentive algorithms.
\begin{itemize}
        \item \textbf{Aggregation Algorithms:} These algorithms primarily address how to effectively aggregate local models from different participants to form a global model. The most common algorithm is FedAvg \cite{mcmahan2017communication}, where participants upload the weights of their local models to the blockchain network in each round of training. The global model is then acquired by calculating the average of these weights through smart contracts. Additionally, there are more sophisticated aggregation algorithms, such as FedProx \cite{li2020federated}, SCAFFOLD \cite{karimireddy2020scaffold}, FedPD \cite{Zhang2020FedPD}, and FedBN \cite{li2021fedbn}. 
        In addition, by constructing an asynchronous FL system \cite{liu2023aedfl,liu2023fedasmu}, it is possible to address the security challenges posed by centralized models, achieving privacy, fault tolerance, and reliable data sharing \cite{wang2022asynchronous}. Based on this architecture, a FL asynchronous aggregation protocol based on permissioned blockchain is proposed that can effectively alleviate the synchronous FL algorithm \cite{che2023federated,liu2022multi,jin2022accelerated,zhou2022efficient} by integrating the learned model into the blockchain and performing two-order aggregation calculations. 
        
        \item \textbf{Security Algorithms:} These algorithms are dedicated to conducting model training and aggregation while preserving the privacy of the participants. The most common methods employ encryption technologies like homomorphic encryption and secure multi-party computation to ensure the weights of the participants' models are not leaked during the upload and aggregation process. Some methods leverage characteristics of blockchain, such as decentralization and immutability, to enhance the system's resistance against attacks.

\begin{itemize}
    \item \textbf{Secure Multi-Party Computation (SMPC):}
    Secure Multi-Party Computation (SMPC) is a computational model for protecting data privacy, allowing multiple parties to compute without disclosing their private data. In SMPC, each party holds a portion of private data, and computations can be performed on encrypted data to maintain privacy.
    
    In SMPC, parties communicate and interact via protocols, collectively calculating the final result without having to directly expose their private data. SMPC often involves the use of encryption techniques, cryptographic protocols, and distributed algorithms to ensure data privacy and security. During computation, parties can use technologies such as homomorphic encryption, secret sharing, zero-knowledge proofs, and secure multi-party computation to ensure the security of the computation process and results.
    \item \textbf{Homomorphic Encryption (HE):}
    HE is an encryption algorithm that satisfies the properties of homomorphic operations on ciphertexts. That is, after the data undergoes homomorphic encryption, a specific calculation is performed on the ciphertext, and the result of the ciphertext calculation, after corresponding homomorphic decryption, is equivalent to the same calculation performed directly on the plaintext data. This realizes a state of "computable but invisible" for the data. By using homomorphic encryption technology, computations can be carried out on ciphertexts without the need for a key, which not only reduces communication costs but also balances the computational costs among all parties.

    \item \textbf{Differential Privacy (DP):}
    Differential privacy is a technology that protects the underlying user privacy information in data by adding disruptive noise. The principle ensures that even if an attacker has mastered all other information except for one piece, they still cannot infer that piece of information. The most common method is to add noise conforming to a certain distribution to the result, randomizing the query result. The main issue to address in differential privacy is data utility. Since it is necessary to incorporate randomness into the query result, it could potentially lead to a decrease in data usability \cite{li2021local}.
    
    \item 
\end{itemize}

        \item \textbf{Optimization Algorithms:} These algorithms aim to optimize the performance of FL, such as reducing the number of training rounds, communication overhead, and enhancing model accuracy. The most common methods include gradient compression and gradient pruning to reduce communication overhead. Other techniques utilize asynchronous and local updates to decrease the number of training rounds.

        \item \textbf{Incentive Algorithms:} These algorithms consider how to motivate participants to join in FL and distribute incentives fairly. Although FL has shown great advantages in enabling collaborative learning while protecting data privacy, it still faces an open challenge of incentivizing people to join the FL by contributing their computation power and data. A Deep Reinforcement Learning (DRL) based incentive mechanism as a solution addresses unique challenges of unshared information and contribution evaluation difficulties in FL \cite{zhan2020learning}. Efficiency of this mechanism is demonstrated via numerical experiments, compared with baseline approaches.
        On the other hand, cross-disciplinary areas such as economics and game theory are also discussed in the context of designing incentive mechanisms for FL \cite{tu2022incentive}.
        The paper elucidates various economic and game models, aiming to understand the motivations behind their use in FL incentive mechanism design. It provides detailed reviews, analyses, and comparisons of different economic and game theoretic approaches for designing a variety of FL incentive mechanisms.

\begin{itemize}
    \item \textbf{Reputation based FL:} 
    To address existing issues in federated learning, such as redundant transmissions, network congestion, as well as security and privacy concerns, reputation-based approaches have been proposed. Typically, reputation-based methods involve measuring the reliability of participants by designing a security mechanism. The blockchain-based reputation system \cite{zhao2020mobile}, which increments the reputation value of clients contributing correct and useful model parameters and decrements for those uploading malicious parameters, influencing client selection for subsequent training rounds. A reputation mechanism framework RepBFL \cite{chen2021repbfl}, merging blockchain and FL for applications in the Internet of Vehicles (IoV). By leveraging blockchain, it ensures shared data protection and the selection of high-reputation nodes for FL, alongside evaluating the reliability of vehicles in IoV. The approach presented in \cite{zhang2021blockchain} employs a reputation-based evaluation using model quality parameters and blockchain to gauge worker reliability and maintain reputation values. In \cite{lyu2020collaborative}, research on collaborative fairness in FL leads to the development of a collaborative fair FL framework, CFFL. This introduces a reputation mechanism based on empirical individual model performance, mediating participant rewards to maintain fairness across communication rounds. Lastly, \cite{kang2019incentive} regards reputation as a metric quantifying the reliability and trustworthiness of mobile devices. Its multi-weight subjective logic model is employed for reputation calculation, with consortium blockchain technology securing reputation storage in a decentralized manner. Notably, the reputation calculation involves the task requester selecting eligible worker candidates based on resource information. The reputation value of candidate workers is then computed based on direct reputation feedback from interaction history and indirect reputation feedback from other task requesters, all of which is stored and managed on an open-access reputation blockchain.
    
    \item \textbf{Payment based FL:}
    FedCoin \cite{liu2020fedcoin} employs Shapley Values (SVs) for a feasible SV-based profit distribution that equitably mirrors contributions to the global FL model. Here, the blockchain consensus entities deploy the Shapley Proof-of-Stake protocol (PoSap) for the calculation of SVs and creation of new blocks. 
    Constructing an FL protocol on a public blockchain network can resolve the challenges related to monitoring worker behavior and guaranteeing protocol adherence  \cite{toyoda2020blockchain}. This protocol embeds competition into BCFL, rewarding only those workers whose contributions are valuable and naturally discouraging deviation from the protocol. The mobile-crowd FL system \cite{jiang2022reward} incentivizes mobile devices to train accurate models by offering rewards based on individual contributions. A Stackelberg game models interactions between the server and devices, and two reward policies, namely, the size-based and accuracy-based policies, are compared under different definitions of individual contribution.
    In addition, the challenge of transparently assessing contributions from different data owners in a cross-silo horizontal federated learning setup is tackled by quantifying data owners' SV-based contributions with adjustable precision, safeguarding their privacy \cite{ma2021transparent}.

    \item \textbf{DeepChain:}
    DeepChain \cite{weng2019deepchain} is a collaborative framework for training deep learning models with joint participation from clients. It guarantees data confidentiality, computational verifiability, and offers incentives to participants. The incentive mechanism in DeepChain is orchestrated around timeout checks and monetary penalties, fostering fairness among participants. It takes punitive measures in scenarios where participants fail to meet deadlines or inaccurately execute functions, by imposing monetary penalties, confiscating the pre-deposited funds from dishonest participants and redistributing them among honest ones.
    
    \item \textbf{Mechanism Design:}
    Designing a mechanism is the objective for achieving effective incentives. The survey \cite{zeng2021comprehensive} meticulously explores incentive mechanisms for federated learning. It compiles existing incentive mechanisms and categorizes them based on key techniques, such as the Stackelberg game, auction, contract theory, Shapley value, reinforcement learning, and blockchain.
    Furthermore, to address the challenges posed by malicious participants in large-scale collaborations, Refiner is proposed \cite{zhang2021refiner}. The system resides on the Ethereum public blockchain platform and operates an incentive mechanism rewarding participants based on the volume of their training data and the performance of local updates. For dealing with malicious actors, Refiner deploys a committee of randomly selected validators. These validators penalize unscrupulous participants by denying rewards and eliminating corrupt updates from the global model.
    
\end{itemize}

\end{itemize}

In BCFL, there is a need to craft an effective incentive mechanism that stimulates active participation among participants and recognizes their contributions. In addition, privacy-preserving methods should be employed to protect data privacy of participants throughout the FL process. Subsequently, its training process should be optimized for the system to achieve the best performance. By addressing the above issues, BCFL systems can motivate participants while ensuring data privacy and security.

\subsection{Blockchain Consensus}

In the context of FL system implementation, consensus pertains to the agreement or protocol employed by participants to synchronize their models and collectively make decisions. Consensus algorithms, such as PoW or PoS, are commonly utilized in blockchain-based systems to establish agreement among distributed participants. These algorithms ensure that all participants agree on the validity of model updates and prevent malicious actors from tampering with the system. 

The consensus mechanism is to complete the verification and confirmation of transactions in a very short period of time through the voting of special nodes. If nodes with disparate interests can reach a consensus on a transaction, it implies a broader network consensus. With the evolution of blockchain technology, the term consensus mechanism has become widely recognized, and various innovative consensus mechanisms continue to emerge.

Consensus holds paramount importance in blockchain technology as it safeguards the integrity, security, and immutability of the distributed ledger. Numerous consensus algorithms are employed in blockchain networks, each characterized by its unique attributes and trade-offs.

\begin{itemize}
    \item \textbf{Proof of Work (PoW):}
    Bitcoin uses the PoW(Proof of Work) workload proof mechanism, and later Ethereum is the PoW and PoS(Proof of Stake) consensus mechanism. PoW is the equivalent of figuring out a difficult math problem that gets harder over time. Although PoW is a consensus mechanism recognized by everyone, computing consumes a lot of energy and may indirectly affect carbon emissions and the environment.
    BCFL \cite{qu2020decentralized}, a distributed hash table for efficient block generation is introduced. This solution employs a proof-of-work consensus mechanism to ensure consistency in the global model.
    LearningChain \cite{chen2018machine}, a decentralized federated system leveraging a Byzantine fault-tolerant aggregation algorithm known as l-nearest aggression. The system is based on the PoW consensus, where the leader is selected through competition and the l-nearest algorithm is used to aggregate the gradient. 
    Swarm Learning (SL) \cite{warnat2021swarm} a decentralized machine learning approach combining edge computing and blockchain-based P2P networking. With Swarm Learning, data and parameters are kept at the edge, thereby eliminating the need for a central coordinator.

    \item \textbf{Proof of Stake (PoS):}
    PoS is seen as a more environmentally friendly alternative to PoW. Instead of miners competing to solve problems, validators are chosen to create new blocks based on the amount of cryptocurrency they hold and are willing to "stake" as collateral.
    \item \textbf{Delegated Proof of Stake (DPoS):}
    In DPoS, stakeholders elect a certain number of delegates who validate transactions and create blocks. This method is designed to be more democratic and efficient than traditional PoS.
    \item \textbf{Proof of Training Quality (PoQ):}
    The existing consensus mechanisms, such as Proof of Work (PoW), consume significant computational and communication resources or have limited additional contributions to data sharing. To address this problem, \cite{lu2019blockchain} proposes a consensus mechanism called PoQ has been proposesd that combines FL with differential privacy. PoQ integrates data model training with the consensus process, replacing the meaningless computational work of finding random numbers in PoW with the authentication of model parameter accuracy. 
    
    \item \textbf{Byzantine Fault Tolerance (BFT):}
    This consensus mechanism aims to withstand 'Byzantine' faults, where components may fail and there is imperfect information on whether a component is failed or not. There are several BFT, such as Practical Byzantine Fault Tolerance (PBFT) used in Hyperledger Fabric \cite{2018hyperledger,sun2021permissioned, chen2020methodology}, and the Federated Byzantine Agreement (FBA) used in Stellar \cite{mittal2021Stellar}.

    \item \textbf{Proof of Federation (PoF):}
    Biscotti \cite{shayan2020biscotti}, which combines PoF with consistent hashing and Verifiable Random Functions (VRF) to select critical roles for peer nodes. These roles aid in coordinating the privacy and security of model updates. To prevent peers from poisoning the model through Multi-Krum defense, Biscotti employs differentiated private noise to provide privacy. It also utilizes Shamir secret sharing for secure aggregation. However, when all nodes participate in the consensus, the computational load is too large \cite{ramanan2020baffle}. 

    \item \textbf{RAFT:}
    RAFT is a consensus algorithm used in some permissioned blockchain networks. It elects a leader among a group of nodes, and the leader is responsible for proposing and validating blocks \cite{kim2021improved}. Raft focuses on simplicity and fault tolerance and is designed to be easier to understand and implement than other consensus algorithms.

    \item \textbf{Proof of Federated Training:}
    Proof of Federated Training (PoFT) \cite{chakraborty2022proof}, is a framework for enabling verifiable model training across multiple blockchain networks. It addresses issues such as power consumption/resource wastage in POW and data privacy in blockchain \cite{qu2021proof}.
    In addition, Proof of FL (PoFL) is also employed in vehicular networks, where vehicles compete to become miners by adhering to the FL consensus proof within the blockchain network \cite{ayaz2021blockchain}. Additionally, IPFS and PoFL are utilized to ensure decentralized federated learning security for connected autonomous vehicles \cite{he2021bift}.

    \item \textbf{Committee Consensus:}
    The Blockchain-based FL framework with Committee consensus (BFLC) \cite{li2020blockchain}, utilizes blockchain for global model storage and local model update exchange, eliminating the need for a centralized server \cite{liu2023heterps}. Additionally, an innovative committee consensus mechanism is introduced to reduce the computational load and mitigate malicious attacks.

\end{itemize}

These consensus algorithms all have their strengths and weaknesses and are suited to different use-cases. Selecting the right consensus mechanism is vital for the security, scalability, and efficiency of the blockchain network.
These are some commonly used blockchain consensus algorithms. However, the selection of a consensus algorithm is influenced by factors such as the desired degree of decentralization, security, scalability, and the specific needs of the blockchain network.

\subsection{Application Layer}
FL has various applications across different domains where services based on this technology can be provided.

\begin{itemize}
    
    \item \textbf{Internet of Vehicles:} 
    % \cite{pokhrel2020decentralized} \cite{chai2020hierarchical} \cite{li2020crowdsfl} \cite{billah2022systematic,saraswat2022blockchain,aloqaily2021energy,ayaz2021blockchain,chai2020hierarchical,chen2021repbfl,chen2021bdfl,dibaei2021investigating,he2021bift,liu2021blockchain,lu2020blockchain,otoum2020blockchain} 
    Vehicles are increasingly becoming data generation sources. Data such as GPS location, speed, and road conditions can contribute to better traffic management, route planning, and accident prevention. However, this data is also sensitive \cite{bai2022internet}. BCFL enables aggregation of data from multiple vehicles to train models without sharing the raw data \cite{pokhrel2020decentralized, chai2020hierarchical, chen2021bdfl, otoum2020blockchain, chen2021repbfl, li2022fedhisyn}. Moreover, blockchain can also be used to maintain a tamper-proof record of vehicle interactions and transactions in the network \cite{aloqaily2021energy, lu2020blockchain}.

    \item \textbf{Resource allocation:}
    In large distributed systems, effective resource allocation is critical to maximize efficiency \cite{wang2022incentive}. By applying FL on top of a blockchain network, resources can be allocated dynamically based on the learning from the network usage patterns \cite{li2021blockchain}. In essence, integrating blockchain with FL not only fortifies privacy and reliability but also provides a platform for efficient resource allocation and utilization \cite{deng2021flex}.

    \item \textbf{Edge computing:} BCFL has great potential for applications in edge computing. It provides a secure mechanism for data sharing, resource collaboration and sharing, model updates and upgrades, as well as guarantees for interference resistance and fault tolerance \cite{liu2021blockchain, lu2020blockchain,nguyen2022latency,nguyen2021federated, qu2020decentralized}. By leveraging the computing resources and data of edge devices, BCFL enables intelligent applications to perform inference and decision-making tasks efficiently and securely  \cite{rahmadika2021blockchain,lu2020low,lu2020communication}. It addresses challenges such as data privacy, resource constraints, and unstable environments commonly encountered in edge computing \cite{wan2022privacy, hu2021blockchain}. This approach offers a novel solution for the development and expansion of edge computing scenarios. It has been extensively researched and applied in the Mobile phones scenario, and it is believed that it will provide solutions for the development and expansion of more edge computing scenarios in the future \cite{fan2022mobile, feng2021two, hu2021blockchain,hu2021blockchainB,kong2021achieving}.
    
    \item \textbf{Healthcare:} In healthcare, patient data is sensitive but can be extremely useful for detecting diseases \cite{kumar2021blockchain,passerat2019blockchain, mohammed2023energy} and improving treatment \cite{rahman2020secure,myrzashova2023blockchain}. BCFL allows healthcare institutions to collaborate and learn from a vast amount of patient data without compromising patient privacy \cite{passerat2020blockchain,aich2022protecting,zhang2021blockchain2}. In addition, blockchain can provide traceability of data and computations, increasing the trust in the learned models \cite{wang2021blockchainMedical}.
    % \cite{passerat2020blockchain,singh2022framework,lakhan2022federated} \cite{bhattacharya2019bindaas,jin2021cross,passerat2019blockchain,singh2022framework}
   
    \item \textbf{Energy:} The integration of blockchain with FL presents a transformative approach for diverse energy applications. BCFL can be used to optimize grid operations, facilitates P2P energy trading and sharing across microgrids \cite{bouachir2022federatedgrids}. Moreover, in the context of the Industrial Internet of Things (IIoT), BCFL can address security and privacy concerns associated with credit data sharing in wireless networks \cite{yang2022privacy, otoum2022federated, issa2023blockchain, Huang2023Distance, singh2023fusionfedblock}. It is also being applied for blade icing detection in wind energy turbines, a rapidly growing sector of renewable energy \cite{cheng2022blockchain}.
    
\end{itemize}

\section{Challenges and Open Research Directions}
\label{sec:Challenges}

The amalgamation of blockchain with FL combines the advantages of both realms, establishing a system that prioritizes data privacy and security. Nevertheless, this integration presents its own array of challenges. In this section, we outline these challenges and propose potential research directions.

\subsection{Data Security and Privacy Protection}
In a landscape where FL with blockchain, the paramount concern becomes ensuring the security and privacy of data. Effectively managing sensitive information, such as medical records or personal identities, within a networked environment is a challenge. An in-depth exploration of cryptographic techniques to handle such data without exposing its true content is essential. Future research should delve into new cryptographic solutions tailored to this integrated system.

\subsection{Model Efficiency and Performance Optimization}
Efficient computation for distributed data and models is a fundamental requirement of FL. The core challenges in this domain involve devising algorithms that reduce computational complexity, minimize communication overhead, and enhance the efficiency of model training and inference. Future research endeavors should concentrate on fine-tuning distributed optimization methods, integrating advanced compression strategies \cite{zhang2022fedduap} to alleviate communication burdens, and exploring hardware enhancements for enhanced computational efficiency. An investigation into hybrid models, amalgamating centralized and decentralized training methods, also holds considerable promise.

\subsection{Scalability}
\chen{As blockchain and FL converge towards creating secure, private data management frameworks, scalability emerges as a crucial bottleneck, particularly as the systems scale and demand increases. The integration of off-chain calculations, sidechains, and layer-2 technologies like state channels or Plasma represents a forward-thinking approach to overcoming these challenges. Off-chain calculations offload intensive tasks, reducing main blockchain load, while sidechains manage transactions separately to decrease congestion. Layer-2 technologies, such as state channels or Plasma, facilitate fast transactions atop the existing blockchain, maintaining security. These approaches, leveraging blockchain's and FL's core principles, aim to cultivate scalable, secure, and decentralized learning ecosystems.}

\section{Conclusion}
\label{sec:Conclusion}

As a decentralized and immutable technology, blockchain enables an egalitarian platform for all network participants, mitigates data breaches, and fosters trustworthiness. The combination of FL with blockchain creates a robust learning ecosystem prioritizing data security and user privacy. In this paper, we conduct a thorough exploration of such integrations, providing a comprehensive perspective on FL architectures informed by blockchain. We aim to offer clear insights for researchers and practitioners in this burgeoning field by examining the foundational elements of both realms and highlighting their synergies. We are optimistic that the challenges and research pathways outlined herein will guide the next wave of innovations in decentralized machine learning frameworks.

\bibliographystyle{plain}
\bibliography{reference}

\end{document}